\documentclass[pra,twocolumn,showpacs,preprintnumbers]{revtex4}
\usepackage{graphicx}
\usepackage{dcolumn}
\usepackage{bm}
\usepackage{amsmath}
\usepackage{latexsym}

\begin{document}

\title{Continuous Measurement Enhanced Self-Trapping of Degenerate Ultra-Cold Atoms in a Double-Well:
 Nonlinear Quantum Zeno Effect  }
\author{WeiDong Li$^{1,2}$, Jie Liu$^{1*}$}
\affiliation{$^{1}$Institute of Applied Physics and Computational
Mathematics, Beijing, 100088, People's Republic of China}
\affiliation{$^{2}$Department of Physics and Institute of
Theoretical Physics, Shanxi University, Taiyuan 030006, China}

\begin{abstract}

In the present paper we investigate the influence of  measurements
on the quantum dynamics of degenerate Bose atoms gases in a
symmetric double-well. We show that  continuous  measurements
enhance asymmetry on the density distribution of the atoms and
broaden the parameter regime for self-trapping. We term this
phenomenon as nonlinear quantum Zeno effect in analog to the
celebrated  Zeno effect in a linear quantum system. Under
discontinuous measurements, the self-trapping due to the atomic
interaction in the degenerate bosons is shown to be destroyed
completely. Underlying physics is revealed and possible
experimental realization is discussed.

\end{abstract}

\maketitle

\section{Introduction}
Double-well system is a paradigm to demonstrate quantum tunnelling
phenomena and attracts much attention from the societies of both
theoreticians and experimentalists since the establishment of
quantum mechanics \cite{grifoni}. For a symmetric double-well, the
amplitude distributions of all  eigenstates are  symmetric, so a
particle is expected to oscillate between the two wells and its
chance to visit each well is even \cite{wanggf}. However, there
exist two effects that can break the above symmetry leading to an
asymmetry on the particle's density distribution, i.e., quantum
measurements
\cite{raizen,pascazio,pascazio1,kurizki,kurizki1,david,micheal1,micheal2}
and collisions between particles (many-body effect)
\cite{smerzi,liu,gati01,gati02,gati03}. In the former case, the
quantum measurement couples  the system to outer environment and
induces the quantum de-coherence of the system. As a result, the
quantum tunnelling between  two wells is suppressed completely so
that the particle keeps staying  in one well and has no chance to
visit the other well. This is the celebrated quantum Zeno effect
(QZE) \cite{micheal1,micheal2}. In the latter case, the two-body
collisions among  the degenerate Bose-Einstein condensates (BECs)
atoms  leads to a nonlinear excitation, manifesting a highly
asymmetric density profile of the BECs even in a symmetric
double-well.  This somehow counterintuitive phenomenon has been
observed  recently in labs \cite{gati01,gati02,gati03}.

In the present paper, what we concern is how a measurement affects
the dynamics of the  many-body quantum system characterized by
degeneracy and diluteness. Under a mean-field approximation and
without measurement, the dynamics of this system is described by a
nonlinear Schr\"odinger equation, known as Gross-Pitaevskii
equation (GPE), where the nonlinearity arises from the interaction
between the degenerated atoms and is responsible for the unusual
self-trapping phenomenon. Considering the measurement carried out
by  a position meter, the GPE is modified to be a stochastic
nonlinear differential equation. With solving the stochastic
equation we  achieve insight into how the measurement affects the
self-trapping of BECs. Our main result is that continuous position
measurements enhance asymmetry on the distribution of the atoms
density profile and broaden the parameter regime for
self-trapping. We term this phenomenon as nonlinear quantum Zeno
effect. However under discontinuous measurements, the
self-trapping phenomenon could be destroyed. Physics behind the
above phenomena is revealed and possible experimental realization
is discussed.

This paper is organized as follows: We first briefly introduce the
quantum measurement theory  in Sec. II. In the third section, we
describe our quasi 1-D double-well model of BECs. In Sec.IV, we
discuss the influences of the continuous or discontinuous
measurements on the dynamics of the system. Finally, we summarize
and discuss our work in section V.

\section{measurement theory}
In the physics community, the measurement side of quantum theory
is one of the fundamental issues \cite{heisenberg,omna,zurek}.
There exist various theories about this topic depending on
concrete physical systems and measured physical quantities
\cite{fearn,schenzle}. For the BECs system we consider, we make
use of continuous (in time) position measurement theory
\cite{micheal1,micheal2}. Within the framework of this theory, the
position and the momentum of the system and the meter are denoted
by $ \{x(t), p(t)\}$ and $\{X(t), P(t) \}$, respectively. Usually
we can describe the pseudo-classical meter as $\langle Q|X\left(
t\right) ,P\left( t\right) \rangle =\left( 2\pi \sigma ^{2}\right)
^{-1/2}\exp \left\{ -\frac{\left[ Q-X\left( t\right) \right]
^{2}}{4\sigma ^{2}}\right\} \exp \left\{ \frac{i}{\hbar }QP\left(
t\right) \right\}$ in the $Q$ (position of the meter)
representation, where $\sigma$ denotes the uncertainty of the
meter pointer. Considering that the meter is pseudo-classical, one
can treat this state as $|X\left( t\right) ,0\rangle $
\begin{equation}
|X\left( t\right) ,0\rangle =\left( 2\pi \sigma ^{2}\right)
^{-1/2}\exp \left\{ -\frac{\left[ Q-X\left( t\right) \right]
^{2}}{4\sigma ^{2}}\right\}. \label{mete}
\end{equation}
This is one of the features of this theory that the higher up the
chain from the system to the observer the cut is placed, the more
accurate the model of the measurement will be. The interaction
between the system and the meter can be described as
\begin{equation}
\hat{H}_{\text{SM}}=\gamma \hat{P}\left[ \hat{x}-\langle x
\rangle_s \right] , \label{int}
\end{equation}
where $\langle x \rangle_s$ is the average position of the system
and $\gamma$ represents the relaxation rate of the meter pointer
to the system mean position $\langle x \rangle_s$. This
interaction translates the position of the meter by an amount
proportional to the average position of the system. The state of
the system-meter at time $t$ is taken to be
\[
|\Phi \left( t\right) \rangle =|X\left( t\right) ,0\rangle \otimes
|\Psi \left( t\right) \rangle ,
\]%
where $|\Psi \left( t\right) \rangle $ is a system state vector.
The combined state at time $t+\tau $ is
\begin{equation}
|\Phi \left( t+\tau \right) \rangle =\exp \left\{ -\frac{i\gamma
\tau }{ \hbar }\hat{P}\left[ \hat{x}-\langle x \rangle_s \right]
\right\} |X\left( t\right) ,0\rangle \otimes |\Psi _{0}\left(
t+\tau \right) \rangle , \label{eequ}
\end{equation}
where
\begin{equation}
|\Psi _{0}\left( t+\tau \right) \rangle =exp \left ( - \frac{i
\gamma \tau}{\hbar} \right ) |\Psi _{0}\left( t\right) \rangle .
\label{eq1}
\end{equation}
From Eq.(\ref{eequ}) and Eq.(\ref{eq1}), we can see that the
evolution of the combined system-meter is purely unitary. The
meter has undergone the desired translation because of the
interaction with the system. Meanwhile the system has undergone
its usual free evolution combined with a measurement back-action
from the meter pointer.

After considering the read-out, the evolution of the combined
equations in the first order of $\tau $ can be written as a
stochastic differential equation
\begin{eqnarray}
\frac{d}{dt}|\Psi \left( t\right) \rangle &=&(-\frac{i}{\hbar
}H_{0}- \frac{\Gamma }{2}\left[ x-\langle \hat{x}\rangle
_{s}\right] ^{2}  \nonumber
\\
&&+\sqrt{\Gamma }\xi \left( t\right) \left[ x-\langle
\hat{x}\rangle _{s} \right] )|\Psi \left( t\right) \rangle
\label{sde}
\end{eqnarray}
where $\Gamma=\gamma^2 \tau / 8 \sigma^2$ denotes the interaction
strength between the meter and the system. The noise term $\xi
\left( t\right) $ indicates Gaussian noise of standard deviation
$1$ and can be modelled by white noise $\langle \xi \left(
t\right) \xi \left( t^{\prime }\right) \rangle =\delta \left(
t-t^{\prime }\right) $.

The above  stochastic Schr\"odinger equation  is equivalent to a
stochastic master equation for the selective evolution of the
conditioned density operator \cite{micheal1}. It is conditioned on
the entire history of the meter readout X(t). If we were only
interested in the non-selective evolution of the system, then we
would have to discard all knowledge of the evolution of the
system. This is achieved in the usual manner of averaging over all
possible meter readouts at all times t. In our case this simply
amounts to averaging over the stochastic term in Eq.(5), which
gives zero.  The process of photon scattering on a BEC that
results in population difference measurement can be modelled by
the non-selective evolution of the density operator \cite{huang}.
In our following discussions, for generality, we include both the
de-coherence term and the noise term modelling the nondestructive
measurements on BECs.

\section{Double-well Model}
\begin{figure}[tbp]
\includegraphics[width=3.2 in]{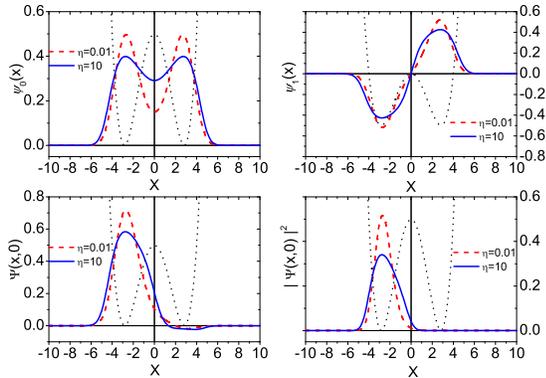}
\vspace{-0.0 in} \centering \caption{(Color online) The top two
pictures show ground state and the first excited state,
respectively. The bottom two show the wave function and the
density profile of the initial sate, respectively. } \label{fig1}
\end{figure}

We will focus on the case of BECs trapped in a quasi-one
dimensional symmetrical double-well. Along the lateral directions
the BECs is tightly confined so that the  Hamiltonian governing
the dynamics of BECs  reduces to the following 1D form,
\begin{equation}
\hat{H}_{0}=-\frac{1}{2}\frac{\partial ^{2}}{\partial x^{2}}+V\left(
x\right) +\eta |\psi \left( x\right) |^{2},  \label{hamilt}
\end{equation}
after re-scaling the energy unit by $\frac{\hbar ^{2}}{mL^{2}}$
and the length unit by $L$ (effective size of the double-well
potential); $m$ is the mass of the trapped atom and $\eta$ is the
nonlinear parameter that is proportional to the 1-D reduced s-wave
scattering length between the degenerate atoms and the total
number of the atoms. As the size of the two well potential
realized in the experiment \cite{gati01} is around $L \sim 13\mu
m$ and the mass of the alkali atom is around $m \sim 10^{-25}kg$,
the order of the time in our paper should be
\[
\frac{mL^{2}}{\hbar }\sim \frac{10^{-25}\times
10^{-12}}{10^{-34}}\sim 10^{-3}s.
\]
$200$ time duration in our dimensionless unit  corresponds to
$0.2s$, that  is within the lifetime of the BECs under the present
experimental conditions. So, our calculations are extended up to
time moment of  $200$ in  following. Meanwhile, the energy scale
is around khz. The double-well potential is expressed by
\begin{equation}
V\left( x\right) =-\frac{1}{4}x^{2}+\frac{1}{64}x^{4}.
\end{equation}
To investigate the tunnelling dynamics of BECs in this
double-well, we start from  a superposition of ground and the
first excited state of stationary GPE, i.e., $\hat{H}_0 \psi_{0,1}
(x) = \mu_{0,1} \psi_{0,1} (x)$, where $\psi_{0,1} (x)$ are ground
state and the first excited wave function respectively and the
corresponding chemical potential are $\mu_{0,1}$. The coefficients
of the superposition are chosen so that the BECs are localized in
one well initially, i.e., $t=0$, as shown in Fig.\ref{fig1}.
\begin{equation}
|\Psi \left(x, 0\right) \rangle =\frac{1}{\sqrt{2}}\left( \psi_{0}
(x) \pm \psi_{1} (x) \right) .  \label{ini}
\end{equation}
We should mention that, in addition to the above symmetric states
having  linear counterpart, the GPE allows eigenstates that are
totally asymmetric, so called symmetry-breaking states, in the
strong interaction case.  These asymmetric eigenstates are crucial
in understanding the dynamics of BECs and are the source of
unusual self-trapping phenomenon \cite{kutz01,liunew}. Our initial
state is chosen as a superposition of ground and the first excited
states and has asymmetric property, but it has nothing to do with
the above asymmetric eigenstates. And further, because the
superposition principle breaks down in the nonlinear case
\cite{liunew}, we cannot predict  the temporal evolution  of a
superposition state from the superposition principle.

In the absence of the nonlinear interaction between atoms, i.e.,
$\eta=0$, our Hamiltonian (\ref{hamilt}) reduces to one depicting
the motion of  single particle in double-well potential and the
QZE is observed in the presence of continuous measurements
\cite{micheal1,micheal2}. When the nonlinear interaction is
present and larger than a critical value $\eta_c$
\cite{wanggf,smerzi,liu}, the somehow counterintuitive
self-trapping phenomenon occurs even without performing
measurement. In the following part of this section, the above
interesting behavior will be   demonstrated. Under the combined
nonlinear interaction and quantum  measurements, the dynamics of
BECs is  dramatically  influenced and  detailed discussions will
be presented in next section.

\begin{figure}[tbp]
\includegraphics[width=3.2 in]{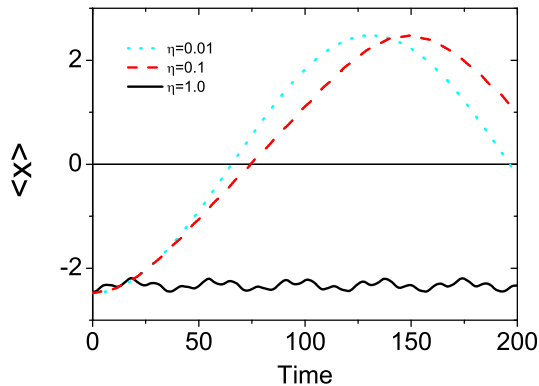}
\vspace{-0.0 in} \centering \caption{(Color online) The Josephson
Oscillation and Self-trapping effect at $\Gamma =0$ } \label{fig2}
\end{figure}

Without performing measurements, i.e., $\Gamma =0$, the dynamics
of the system strongly relies on the interaction between the
particles, the time-dependent  GPE governs the evolution of the
BECs. As in Ref. \cite{liu}, we apply the operator-splitting
method to calculate the evolution of the wave function and then
calculate average position through the formula $ \langle x(t)
\rangle= \int \psi (x,t)^*x \psi (x,t) dx$. In Fig.\ref{fig2}, we
show the dynamic evolution of the coherent superposition state
(\ref{ini}) for different nonlinear interaction parameters $\eta$
by plotting the average position $\langle x(t) \rangle$ with
respect to  time. It shows quite different dynamic behaviors,
depending on the values of the nonlinear interaction. For the weak
interaction case, i.e.,  $\eta < \eta_c \sim 0.143$, the BECs
demonstrate a coherent Josephson oscillation between two wells,
while for $\eta > \eta_c$, the BECs are trapped in one well, a
phenomenon  so called as the self-trapping. In our case, the
population difference between the two wells of our initial state
is quite large, so the self-trapping condition is easier to be
satisfied compared to usual cases \cite{wanggf,smerzi}.

\section{The influence of the measurement}

In this section, we will investigate the influence of the
measurements on the dynamics of the system. According to the
measurement theory, the continuous measurements is modelled by
treating a sequence of $n$ measurement operations described in the
Sec.I, each separated by a time interval  $\Delta t$, over a total
time duration $T=n \Delta t$. The continuous limit is obtained by
taking the limit $\Delta t \rightarrow 0$ \cite{micheal1}. The
probability that the system is found on the initial state at each
independent measurement is $P_0(t=T) \approx 1- 2 (\Delta H)^2
\frac{T^2}{n}$, where $(\Delta H)^2=\langle \psi_0 | H^2| \psi_0
\rangle - \langle \psi_0| H | \psi_0 \rangle^2$.  As is shown in
following, one can recover the QZE in the linear dynamics of the
system ($\eta=0$ in (\ref{hamilt}) ) by setting the time step as
$\Delta t = 10 ^{-3}$, which corresponds to the separated time
$\tau$ between two consecutive  measurements. But if we increase
the time interval to $\Delta t = 10 ^{-2}$, the continuous
measurements condition is broken, we fall into the regime of  the
discontinuous measurements.

\subsection{Continuous measurements: The nonlinear Zeno effect}

With applying the  continuous measurements, the dynamics of the
system is governed by Eqs.(\ref{sde}). It is interesting to note
that the  equation (\ref{sde}) can be split into two parts. One is
the normal free evolution part of the nonlinear Schr\"{o}dinger
equation, the other part is the time-dependent random part. To
solve the dynamic evolution of this problem, we first transform
the wave function into the ``interaction" representation and
integrate the time-dependent part (the second part) with the
Weiner increment method. Then we apply the inverse operation to
obtain the wave-function in the coordinate space. Please note that
the transformation to the ``interaction" representation is
time-independent unitary transformation described as $\exp(-iH_0
\Delta t /\hbar)$, where $\Delta t$ is the time step.

\begin{figure}[tbp]
\includegraphics[width=3.2 in]{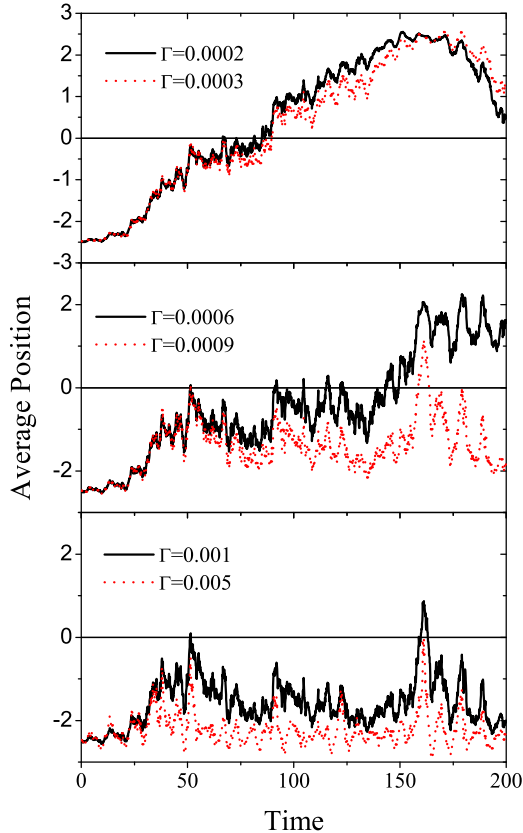}
\vspace{-0.0 in} \centering \caption{(Color online) The average
position of the BECs as the function of time with increase the
value of $\Gamma$ from the top to the bottom in the linear case. }
\label{fig3}
\end{figure}

The Weiner increment $dw$ helps us to numerically integrate
Eqs.(\ref{sde}) with a random term. According to the theory of
Stochastic different equations, we  introduce $dw$ as follows,
\begin{equation}
\frac{dw}{dt}=\xi \left( t\right).
\end{equation}
Considering that our noise term $\xi(t)$ is Gaussian white noise
\cite{evans}, we can write Weiner increment as
\begin{equation}
dw=\sqrt{\tau } \text{ Rand}\left( 0,1\right),
\end{equation}
where $\text{Rand}(0,1)$ denotes the Gaussian random number with
zero mean value and $1$ standard variance.

Besides the time-dependent random noise term, we also have to
include the idea of the quantum trajectory theory
\cite{micheal1,micheal2}. In the quantum trajectory theory, each
measurement operator can be regarded as a ``state preparation"
procedure. This procedure creates an ensemble of systems each
possessing a certain state. Actually, the measurement operator is
realized by projecting the state into one of the position meter
state (\ref{mete}). Physically, we can understand this process as
following: After finishing  each measurement, we have destroyed
the measured quantum state and the state collapses  into one of
the states of meter pointer. So in the next step, we exactly start
our evolution from this quantum state. Therefore, after each
measurement  quantum state is prepared through a normalization
process.

In Fig.\ref{fig3}, we plot the temporal evolution of average
position of the particle for linear case $\eta=0$. It shows that,
for weak measurement (i.e., small $\Gamma$), the particle
oscillates between the two wells, whereas the particle will stay
in the left well (its initial location) with increasing the
interaction strength between the meter and the system. The above
calculation clearly demonstrates the celebrated QZE. In the above
calculation, the time step is set as $\Delta t =10^{-3}$, which is
small enough to guarantee that our measurements are continuous.
Without measurement, the period of the oscillation is around
$T_{os} \sim 250$. From our Eq.(5), the de-coherence term is
estimated proportional to $\Gamma /2 \times (\Delta x_0)^2$, where
$\Delta x_0$ is the distance between two wells and is $2\sqrt{8}$.
So it is expected that when $\Gamma \sim (2 \times \pi)/(16 \times
T_{os}) \sim 0.0015$ the de-coherence effect caused by the
measurement will be significant so that the linear QZE is
observed. The above simple estimation excellently agrees with our
numerical calculation as shown in Fig.3, where we see at $\Gamma >
0.001$ the particle turns to be localized in its initial well.

However, we find that the above simple picture is no longer
available when the nonlinear interaction between atoms emerges. In
this situation, we find that the de-coherence effect caused by the
measurement is significantly enhanced by the atomic interaction.
\begin{figure}[tbp]
\includegraphics[width=3.2 in]{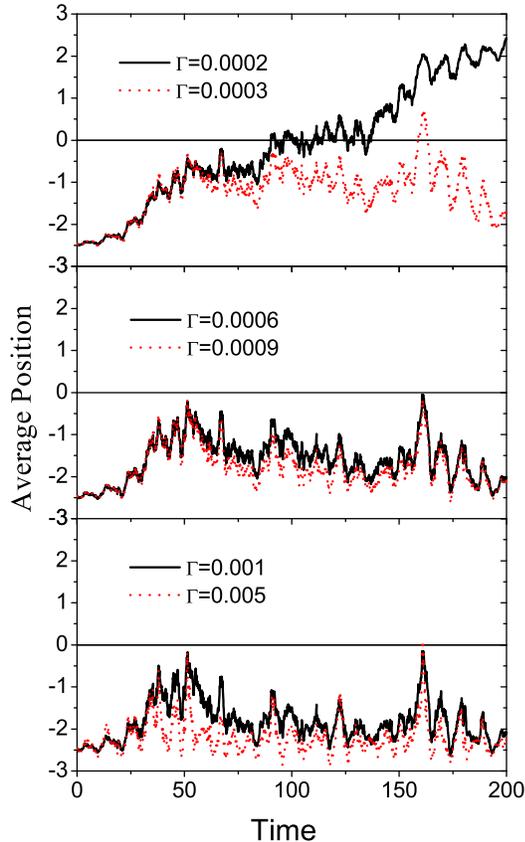}
\vspace{-0.0 in} \centering \caption{(Color online) Same as
Fig.(\ref{fig3}) in the case of $\eta=0.1$} \label{fig4}
\end{figure}

\begin{figure}[tbp]
\includegraphics[width=3.2 in]{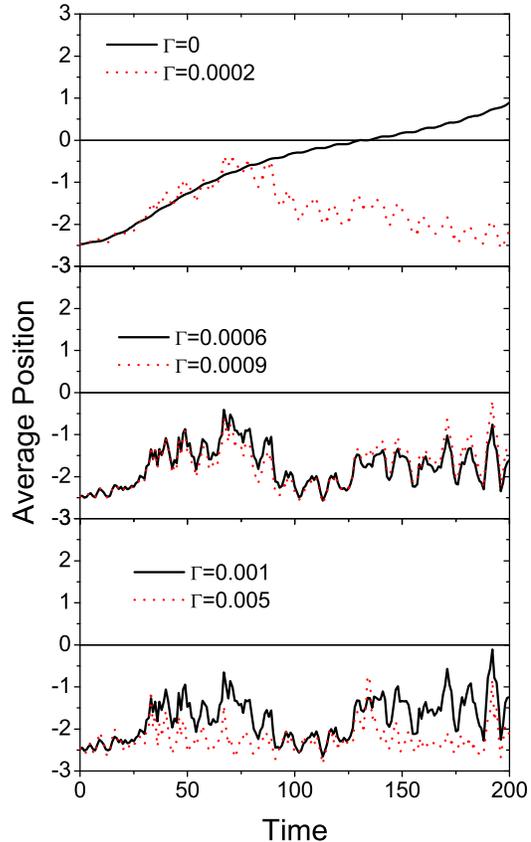}
\vspace{-0.1 in} \centering \caption{(Color online) Same as
Fig.\ref{fig3} in the case of $\eta_c=0.143$} \label{fig5}
\end{figure}

We plot the average position of BECs in the presence of the
nonlinear interaction between atoms, i.e.,  $\eta =0.1$, in
Fig.\ref{fig4}. In this case, the nonlinear interaction is smaller
than the critical nonlinear interaction $\eta_c$, so that the BECs
show a Josephson oscillation  between the two wells in the absent
of measurements $\Gamma =0$ (see Fig.\ref{fig2}). With increasing
the interaction strength between system  and  meter,
Fig.\ref{fig4} shows that the BECs turn  to stay in one well
manifesting that the averaged position is smaller than zero. The
critical value $\Gamma$ for the occurrence of Zeno effect is
modified from $\Gamma=0.001$ in linear case to $\Gamma=0.0003$ in
weak nonlinear case ($\eta=0.1$). In this case the period of the
nonlinear Josephson oscillation is $T_{os} \sim 300$. From the
simple picture given in the above discussions we estimate that the
de-coherence effect become significant only when $\Gamma >
0.0012$. However, at $\Gamma=0.0003$ we already have observed the
blockage of the Josephson oscillation. This analysis implies that
the interaction among BECs atoms significantly enhance the
de-coherence effect. As a result, even a moderately weak nonlinear
interaction can play a dramatic role  and significantly change the
movement of the atoms. So we conclude that the nonlinear
interaction broadens the regime of the localization of atoms  and
makes it easy to observe the quantum Zeno effect. Comparing with
the linear quantum Zeno effect, we term  this phenomenon as
nonlinear quantum Zeno effect. On the other hand, the continuous
measurements suppress the oscillation between the two wells and
enhance the self-trapping by enlarging the regime of
self-trapping. To show this point more clearly, we further plot
the average position of the BECs with $\eta_c =0.143$ in
Fig.\ref{fig5}.

\begin{figure}[tbp]
\includegraphics[width=3.2 in]{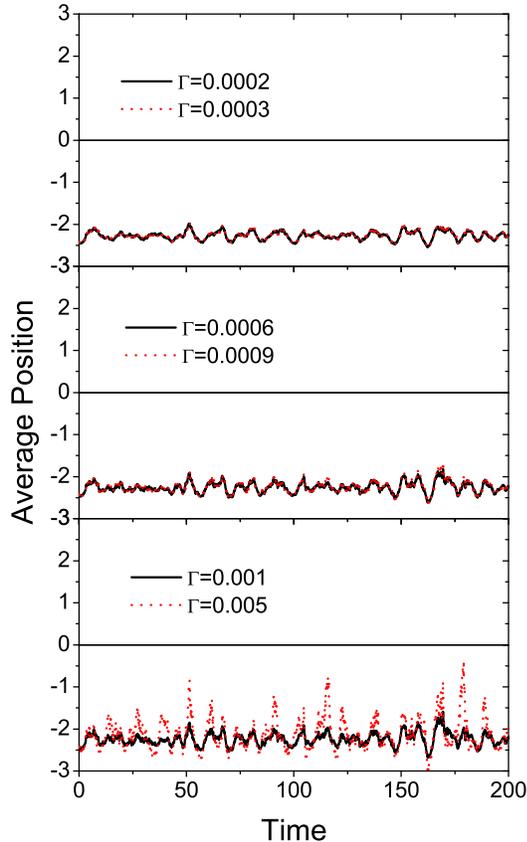}
\vspace{-0.1 in} \centering \caption{(Color online) Same as
Fig.\ref{fig3} in the case of $\eta=1.0$} \label{fig6}
\end{figure}

When the nonlinear interaction is larger than the critical value,
 we also calculate the average position of BECs with respect to
time as shown in Fig.\ref{fig6}. In the presence of  the strong
nonlinear interaction, the BECs atoms are  self-trapped even
without performing the measurement. From Fig.\ref{fig6}, we see
that, the continuous measurements only  cause  small displace of
the averages position of BECs, but can not drive the BECs to
tunnel through the barrier. This means that the continuous
measurements do not destroy this kind of  nonlinear self-trapping.

\begin{figure}[tbp]
\includegraphics[width=3.2 in]{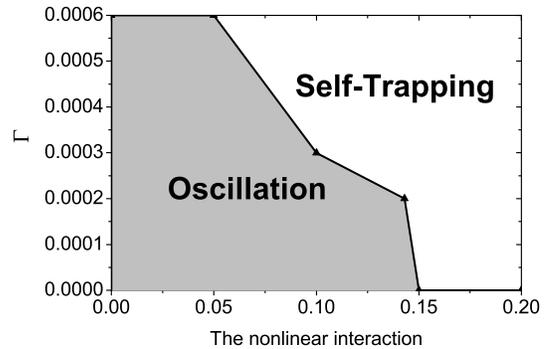}
\vspace{-0.1 in} \centering \caption{The diagram of the parameter
regime of the self-trapping, the light grey region shows the
regime of the coherent oscillation } \label{fig7}
\end{figure}

In Fig.\ref{fig7}, we plot the parameters diagram  of the
occurrence of the self-trapping. We find that, the continuous
measurements broaden the parameter regime for the emergence of
self-trapping. For example, the continuous measurements of
strength $\Gamma=0.0006$ reduces the value of the critical
nonlinear interaction $\eta_c$ from 0.143 to 0.05.  On the other
hand, the nonlinear interaction between the degenerate atoms makes
it easier to observe  QZE  by reducing the critical coupling
strength $\Gamma$ from $0.001$ in linear case (see Fig.\ref{fig3})
to $0.0003$ in the case of $\eta=0.1$ (see Fig.\ref{fig4}).
Therefore, we conclude that, the combination of the continuous
measurements and the nonlinear interaction greatly broaden the
parameter regime for observing QZE and occurrence of the
self-trapping.

\subsection{discontinuous measurement}

So far, we have investigated the effect of the continuous
measurements on dynamics of the nonlinear quantum system. In this
part, we want to know what happen if  our measurement is not
continuous, i.e., to study the influence of the discontinuous
measurements on the self-trapping effect.

\begin{figure}[tbp]
\includegraphics[width=3.2 in]{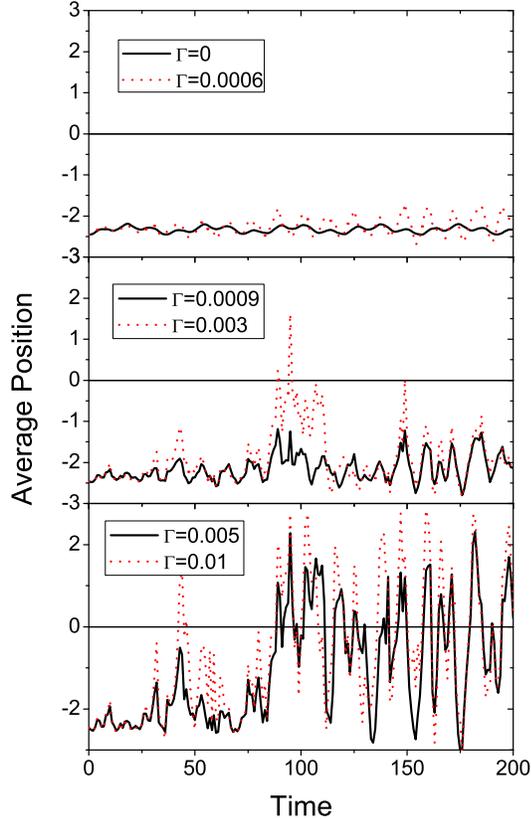}
\vspace{-0.1 in} \centering \caption{(Color online) The average
position as the function of time for time step $\Delta \tau
=10^{-2}$ with the nonlinear interaction is $\eta=0.5$}
\label{fig8}
\end{figure}

To investigate the effect of  discontinuous measurements, we will
modify the time interval of the sequence of the measurement from
$10^{-3}$ to $10^{-2}$. Considering the barrier height, we can not
increase the parameter $\Gamma$ to too large. But if we keep
$\Gamma=0$ and increase the nonlinear interaction to $\eta=0.5$,
the dynamics of BECs initially localized in one well demonstrates
 a self-trapping phenomenon the same that as shown in Fig.\ref{fig2}.
In Fig.\ref{fig8}, we plot our results in the  same way as the
previous figures from the top to the bottom for various values of
$\Gamma$. It shows that the   oscillation amplitude of the
averaged position of BECs  increases  with the increasing
measurement strength $\Gamma$. In the case of $\Gamma=0.005$, the
particles can cross the barrier and tunnel into the other well.
This means that the discontinuous measurements  break the
self-trapping effect that is induced by the strong interaction
between atoms. In the other words, the discontinuous measurements
but still high frequency measurements have tendency to recover the
macroscopic coherence properties of the system. The similar effect
of the measurements process has been pointed out by investigating
the dynamics of the quantum dynamics of the BECs with two wells
including the nondestructive measurements \cite{walls}.

\section{Conclusion }

In conclusions, we have investigated the influence of both
continuous and discontinuous position measurements on the quantum
dynamics of degenerate Bose atoms gases in a symmetric
double-well. We find that, continuous position measurements
enhance the asymmetric distribution of the atoms density profile
and broaden the parameter regime for self-trapping. Moreover, the
de-coherence effect caused by the measurement is significantly
enhanced by the atomic interaction.  The nonlinear interaction
between the degenerate atoms makes it easier to observe QZE by
decreasing the critical coupling strength $\Gamma$, only over
which observing QZE is possible. Therefore, we conclude that, the
combination of the continuous measurements and the nonlinear
interaction greatly broaden the parameter regime for observing QZE
and occurrence of the self-trapping.  On the other hand, we find
that, discontinuous measurements may break the self-trapping. It
implies that the discontinuous position measurements may enhance
the tunnelling of the BECs atoms. In the present experimental
condition, BECs trapped in the symmetric double-well has been
realized with using optical trap technique \cite{gati01} and
nondestructive measurements can be realized by shining a coherent
light beam through BECs \cite{walls}. The measurement of the mean
coordinate is, roughly speaking, equivalent to that of the
inter-well population difference. Therefore, the photon scattering
on BECs in a double-well potential is suggested to give the
population difference measurement \cite{huang}.
 Our predicted phenomena is hopefully observed with present experimental
technique.

\begin{center}
{\bf{ACKNOWLEDGMENTS}}
\end{center}
WL is supported by the NSF of China (No. 10444002), SRF for ROCS,
SEM, SRF for ROCS, Ministry of Personal of China and SRF for ROCS
of Shanxi Province. JL is  supported by National Natural Science
Foundation of China (No.10474008,10445005), Science and Technology
fund of CAEP, the National Fundamental Research Programme of China
under Grant No. 2005CB3724503, the National High Technology
Research and Development Program of China (863 Program)
international cooperation program under Grant No.2004AA1Z1220.  We
gratefully thank Shi-Gang Chen, Shu-Qiang Duan, Li-Bing Fu, and
Guang-Jong Dong  for the stimulating discussions.

\end{document}